\documentstyle[prd,aps,epsfig]{revtex}

\newcommand{\ETAL}{{\it et al.}}  

\newcommand{\BAR}{{\rm b}}
\newcommand{\MAT}{{\rm m}} 

\newcommand{\SCAL}{{\rm S}}

\newcommand{\GROUPE}[2]{{\rm {#1}}(#2)}
\newcommand{\UUNIT}[2]{\;{\rm {#1}}^{#2}}

\newcommand{\STR}{{\scriptscriptstyle \rm S}}
\newcommand{\INF}{{\scriptscriptstyle \rm I}}

\begin{document}
\tighten
\draft

\twocolumn[\hsize\textwidth\columnwidth\hsize\csname
@twocolumnfalse\endcsname

\title{Evidence against or for topological defects in the BOOMERanG
data ?}

\author{F.~R.~Bouchet$^1$, P.~Peter$^1$, A.~Riazuelo$^2$,
  M.~Sakellariadou$^{1,3}$ }

\address{$^1$Institut d'Astrophysique de Paris, 98bis boulevard Arago,
  75014 Paris, France\\ $^2$Universit\'e de Gen\`eve, Service de
  Physique Th\'eorique, 24 Quai Ernest Ansermet, 1211 Gen\`eve 4,
  Switzerland\\ $^3$Section of Astrophysics, Astronomy and Mechanics,
  University of Athens, Panepistimiopolis, 15784 Zografos, Hellas}

\maketitle

\begin{abstract}
  The recently released BOOMERanG data was taken as ``contradicting
  topological defect predictions''.  We show that such a statement is
  partly misleading. Indeed, the presence of a series of acoustic
  peaks is perfectly compatible with a non-negligible topological
  defects contribution. In such a mixed perturbation model (inflation
  and topological defects) for the source of primordial fluctuations,
  the natural prediction is a slightly lower amplitude for the Doppler
  peaks, a feature shared by many other purely inflationary
  models. Thus, for the moment, it seems difficult to rule out these
  models with the current data.
\end{abstract}

\narrowtext
\vspace{0.2cm}
]

For almost two decades, two families of models have been considered
challengers for describing, within the framework of gravitational
instability, the formation of large-scale structure in the universe.
Initial density perturbations can either be induced by quantum
fluctuations of a scalar field during inflation~\cite{inflation}, or
they may be triggered by a class of topological defects~\cite{TD}. The
inflationary paradigm was proposed in order to explain the
shortcomings of the standard Big Bang model, namely the horizon and
flatness problems. Topological defects, on the other hand, can be
formed during symmetry-breaking phase transitions in the early
universe~\cite{kibble}. Moreover they are naturally predicted by many
realistic models of particle physics aiming at describing interactions
at energies much higher than what is currently reachable in
accelerators~\cite{TD}.

The Cosmic Microwave Background (CMB) anisotro\-pies provide a
powerful tool to discriminate among inflation and topological defects.
CMB anisotropies are conveniently characterized by their angular power
spectrum $C_\ell$, the average value of the square of the
coefficients, $| a_{\ell m} |^2$, of a spherical harmonic
decomposition of the measured CMB pattern. This spectrum fully
characterizes a normal distributed field, and $\ell (\ell + 1) C_\ell$
is the logarithmic contribution to the variance of multipoles $\ell$.

On large angular scales ($\ell \lesssim 50$), both families of models
lead to approximately scale-invariant spectra, with however a
different prediction regarding the statistics of the induced
perturbations. Provided the quantum fields are initially placed in the
vacuum, inflation predicts generically Gaussian fluctuations, whereas
in the case of topological defects models, the induced perturbations
are clearly non-Gaussian, at least at sufficiently high angular
resolution. This is an interesting fingerprint, even though difficult
to test through the data. In the context of inflation, non-Gaussianity
can however also be present, as for example in the case of stochastic
inflation~\cite{stochastic}, or in a class of inflationary models
involving two scalar fields leading to non-Gaussian isothermal
fluctuations with a blue spectrum~\cite{linde1}. In addition, allowing
non-vacuum initial states for the cosmological perturbations of
quantum-mechanical origin, one generically obtains a non-Gaussian
spectrum~\cite{jam}.  Finally, in a cosmological model where
perturbations are induced by inflation with a non-negligible
topological defects contribution, one again expects deviation from
Gaussianity.

Both inflation and defect models predict a spectrum of temperature
anisotropies at the required $10^{- 5}$ level~\cite{COBE} on large
angular scales. However, on intermediate and small angular scales, the
predictions of inflation are quite different than those of topological
defects, due to the different nature of the induced perturbations.
More precisely, the inflationary fluctuations are coherent, in the
sense that the perturbations are initially at the same phase and
subsequently evolve linearly and independently of each other. The
subsequent progressive phase shift between different modes thus
produces the so-called acoustic (or Doppler) peak structure. On the
other hand, in topological defect models, fluctuations are constantly
induced by the sources (defects). Since topological defects evolve in
a non-linear manner, and since the random initial conditions of the
source term in the perturbation equations of a given scale leaks into
other scales, perfect coherence is destroyed~\cite{joao,rd}.

In the inflationary case, coherent adiabatic fluctuations lead to a
rather high first acoustic peak, around $\ell \sim 200$ (for $\Omega
\sim 1$) followed by a set of oscillatory peaks for larger $\ell$. As
a matter of illustration, in the Standard Cold Dark Matter (SCDM)
model with $n_\SCAL = 1$, $h = 0.5$, $\Omega_0 = 1$, $\Omega_\BAR =
0.05$, and vanishing cosmological constant, the second peak is
expected to be three times higher than the low $\ell$ (Sachs-Wolfe)
plateau.

On the other hand, in topological defect models, incoherent
fluctuations lead to a single bump at smaller angular scales (larger
$\ell$). For instance, global $\GROUPE{O}{4}$ textures predict a peak
whose position is shifted to $\ell \simeq 350$ having an amplitude
which is $\sim 1.5$ times higher than the Sachs-Wolfe
plateau~\cite{rm}. Global $\GROUPE{O}{N}$ textures in the large $N$
limit lead to a rather flat spectrum, with a slow decay after $\ell
\sim 100$~\cite{dkm}. Roughly similar results are obtained with other
global $\GROUPE{O}{N}$ defects~\cite{clstrings,num}.  Local cosmic
string predictions are unfortunately not very well established in
detail and range from an almost flat spectrum~\cite{acdkss} to a
single wide bump at $\ell \sim 500$~\cite{mark} with extremely rapidly
decaying tail.  It seems that the microphysics of the string network
plays a crucial role in the height and in the position of the
bump~\cite{rpd,wiggles}.  An interesting point however is that global
textures (and especially global $\GROUPE{O}{N}$ defects in the large
$N$ limit, where every component of the field which produces the
defects evolves almost linearly, see~\cite{dkm}) are in a sense ``less
non-linear'' than local cosmic strings. This translates into the
presence of a very smooth second peak in the spectrum.

Before the first Antarctic BOOMERanG~\cite{BOOMERanG} results, several
experiments had already given some hints for a rise in the power
spectrum at the degree scale ($\ell \sim 200$). Then, more recently,
the TOCO~\cite{toco} experiment and the technical flight over North
America of BOOMERanG itself~\cite{boomprel} detected a maximum in the
spectrum around the same scale, followed by a subsequent decay,
thereby ruling out pure topological defect based scenarios. But
measurements at $\ell > 400$ were rather crude (e.g.,~\cite{CAT}). The
BOOMERanG experiment has  spanned in a single experiment a very
wide range of angular scales, from $\ell \simeq 50$ to $\ell \simeq
800$ and has reached sufficient sensitivity to map out the second and
third peak region of the spectrum.

Since then, the observational situation has dramatically improved.
The analysis of the BOOMERanG-LDB data set, which is exceptional in
data quality, span of scales covered, and angular resolution, firmly
reveals the presence of a relatively narrow peak in the power spectrum
at $\ell \sim 200$ followed by an almost flat plateau, roughly at the
same level as the Sachs-Wolfe part. The first peak is found to be at
the position expected in the case of a flat universe.

The data at higher $\ell$ lead to some confusion. In the first release
of the data~\cite{BOOMERanG}, only a relatively low secondary peak
could be accommodated.  Even though their was no direct measurement of
the height of the second peak, one could still, safely enough, remark
that the relative amplitude between the two peaks was rather high.
Indeed it was well above its expected value, even in the case of a
high baryonic density universe which tends to enhance the contrast
between odd and even peaks (see~\cite{hu} for a detailed description
of the influence of cosmological parameters on the CMB anisotropy
spectrum).  This finding has led to new directions for the explanation
of the CMB anisotropies. We briefly discuss some suggestions below.
Within the context of inflationary perturbations, the data
required~\cite{white} a tilted model with a red spectrum (neglecting
the possible contribution of tensor modes, i.e., gravitational waves,
the measured spectral index would appear to be of order $n_\SCAL
\simeq 0.85$) and/or a very high baryonic density, more than one
standard deviation away from the nucleosynthesis
constraint~\cite{nucleo}. Such conclusions put the simplest
inflationary scenario in difficulty.  Other scenarios were then
proposed in order to explain a relatively low second peak together
with a high and narrow first one, such as, e.g., including a leptonic
asymmetry~\cite{jul}.  Another possibility was to add power at low
$\ell$.  There are several possibilities to do so, for example by
adding a tensor contribution (which is rather flat before $\ell \simeq
100$ and rapidly decays afterwards; note that if one considers an
inflationary scenario with a red spectral index, there is necessarily
such a non negligible tensor contribution), or isocurvature
fluctuations~\cite{enqvist,rl}. These possibilities were discussed in
Ref.~\cite{teg}.

The more recent BOOMERanG data~\cite{BOOMnew} somewhat changed this
picture.  A new estimation of the beam size and of the calibration
uncertainties, led to a significant increase of the power spectrum in
the region of the secondary peaks, as well as a confirmation of the
long expected acoustic oscillations in the spectrum.  Then, the
estimation of the cosmological parameters from the shape of the
spectrum led to values definitely more compatible with the currently
ones derived by many independent astrophysical measurements.  Although
there are still some differences with the new MAXIMA
data~\cite{MAXnew}, this might be seen as the long-awaited advent of
``consistent cosmology''\cite{tegnew}. However, in order to give firm
bounds on the value of the cosmological parameters, we still must
ensure that we have explored the correct region of the parameter
space.  Since there already exist very good fits to the data, this
reduces to ask at what extent the currently favored set of
cosmological parameters might be degenerate with other yet non
explored but physically relevant parameters.

In this article, we illustrate a new degeneracy arising in the data,
that would be due to a small but significant contribution of
topological defects. These mixed perturbation models (i.e., both of
inflationary type and seeded by topological defects) have already been
proposed on other grounds.  Indeed, the formation of topological
defects provides the necessary mechanism in order to successfully exit
the inflationary era~\cite{hybrid} in a number of particle physics
motivated inflationary~\cite{linde2,riotto1,realistic} models.  Within
the context of hybrid inflation~\cite{hybridinfl} in supergravity, a
model leading to a scenario where CMB anisotropies may be produced by
cosmic strings on a $\sim 10^{16} \UUNIT{GeV}{}$ mass scale, and
galaxy formation may be due to inflationary perturbations, was
proposed in Ref.~\cite{riotto1} and further investigated in
Refs.~\cite{markjoao,battye}. In addition, in most classes of
superstring compactification involving the spontaneous breaking of a
pseudo-anomalous $\GROUPE{U}{1}$ gauge symmetry~\cite{anomalous},
strings of the cosmic kind~\cite{anomstring} are also
formed~\cite{globalsuper} and turn out to be local~\cite{localsuper}.
We therefore consider a model in which a network of cosmic strings
evolved~\cite{BB} independently of any pre-existing fluctuation
background generated by a standard cold dark matter with a non-zero
cosmological constant ($\Lambda$CDM) inflationary phase.  As we shall
restrict our attention to the angular spectrum, we can remain in the
linear regime all along our analysis, so that setting $C^\INF_\ell$
and $C^\STR_\ell$ the (COBE normalized) Legendre coefficients due to
adiabatic inflation fluctuations and those stemming from the string
network respectively, we have
\begin{equation}
C_\ell =   \alpha     C^\INF_\ell
         + (1-\alpha) C^\STR_\ell.
\end{equation}
In the above equation, the $C_\ell$ are to be compared with the
observed data and the coefficient $\alpha$ is a free parameter giving
the relative amplitude for the two contributions. In this preliminary
work, we do not vary $C^\STR_\ell$ characteristics and simply use the
model of Ref.~\cite{num}.  Strictly speaking, the anisotropy power
spectrum reported in Ref.~\cite{num} concerns theories of global
defects. However, the finding that vector and tensor modes provide
substantial contribution to the large angular scale anisotropies, hold
for both global defect models~\cite{clstrings,num} and cosmic string
scenarios~\cite{acdkss}.  Moreover, models of global strings as well
as of local strings, are characterized by incoherent perturbations.
We thus believe that the spectrum of Ref.~\cite{num} exemplifies the
power spectra of both global and local cosmic strings. In addition,
another advantage of this model, is that it takes into account the
expansion of the universe. To obtain the power spectrum from numerical
simulations with cosmic strings, one must take into account the
``three-scale model''~\cite{3scale} of cosmic string networks, the
small-scale structure (wiggliness)~\cite{wiggles} of the strings, the
microphysics of the string network~\cite{rpd}, as well as the
expansion of the universe. This is a rather difficult task and to our
knowledge no currently available simulation includes all of them.
\vskip 0.5cm

\begin{figure}
  \centering \epsfig{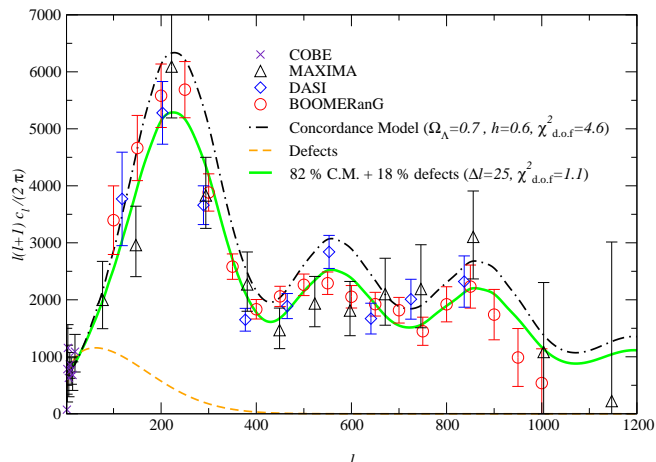}
  \caption{$\ell (\ell + 1) C_\ell$ versus $\ell$ for three different
    models. The upper dot-dashed line represents the prediction of a
    $\Lambda$CDM model, with cosmological parameters set as $n_\SCAL =
    1$, $\Omega_\Lambda = 0.7$, $\Omega_\MAT = 0.3$, $\Omega_\BAR =
    0.05$ and $h = 0.6$ in agreement with all other data but CMB's.
    The lower dashed line is a typical string spectrum.  Both of these
    are seen not to fit the new BOOMERanG, MAXIMA and DASI data
    (circles, triangles and diamonds respectively) and are normalized
    at the COBE scale (crosses).  Combining both curves with the
    extra-parameter $\alpha$ produces the solid curve, with a $\chi^2$
    per degree of freedom slightly above unity. The string
    contribution turns out to be some 18\% of the total. With the
    former BOOMERanG data which produced a much lower second peak, the
    string content was raised to 38\% of the total.}
\label{fig1}
\end{figure}
Figure~\ref{fig1} shows the two uncorrelated spectra as a function of
$\ell$, both normalized on the COBE data, together with the weighted
sum. The best fit, having $\alpha \sim 0.82$ yields a contribution in
strings which is not negligible, although the inflation-produced
perturbations definitely represent the dominant part for this
spectrum. Also, our conclusions are not significantly affected if we
remove the new MAXIMA~\cite{MAXnew} and DASI~\cite{DASI} data. The
fact that the string content extracted from the data is not very large
is of course related to the fact that our choice of value for the
cosmological parameters are very similar to the best fit found by the
BOOMERanG team~\cite{BOOMnew}. For example, the string content would
be larger should we consider the former BOOMERanG
2000~\cite{BOOMERanG} data instead.  Nevertheless, it illustrates that
there is some degree of degeneracy between this model with a string
contribution and the model without strings even with popular
cosmological parameters and ``good'' data. It thus seems to us
necessary to add the string contribution as a new parameter to the
standard parameter space.  Actually, it appears quite natural to
enlarge the parameter space as increasingly high quality data begin to
be at our disposal.  In the same vein, there seems to be a growing
agreement on the fact that a tensor contribution has to be taken into
account when trying to fit the data, and that the presence of non
adiabatic initial conditions also deserves to be more carefully
studied.

We would like to briefly comment upon the ``biasing problem'' in
structure formation. Topological defect models have been sometimes
claimed to be ruled out, since they would not lead to large enough
matter density perturbations, once normalized to the COBE data on very
large scales. This normalization fixes the only free parameter of a
given defect model, namely the symmetry breaking scale.  More
precisely, on scales of $100 \,h^{-1} \UUNIT{Mpc}{}$, which are most
probably unaffected by non-linear gravitational evolution, standard
topological defect models, once normalized to COBE, require a bias
factor of $b_{100} \sim 5$, whereas $b_{100}$ is most probably close
to unity. In cosmic string models, the biasing problem is definitely
not that severe, since it is model dependent. More precisely, the
matter power spectrum is very sensitive to the assumptions made about
string decay~\cite{mark,rpd}. Clearly, the ``biasing problem'' may be
cured within mixed perturbation scenarios, in which structure arises
by the combined effect of both adiabatic density perturbations
produced at the end of inflation and isocurvature fluctuations seeded
by topological defects.

In conclusion, we have found that a mixture of inflation and
topological defects can perfectly well accommodate the current CMB
data, with standard values of the cosmological parameters.
We think that this new, unexpected, degeneracy should be carefully
taken into account in the future CMB data analysis.  We are well aware
that the analysis presented in this paper is very coarse. For
instance, the actual shape of the string spectrum which we used is yet
very uncertain and a more detailed study is required. In particular,
it could be parameterized so that a more thorough analysis could be
performed, not only on one single variable $\alpha$, but over the full
cosmological parameter set in order to incorporate all the effects
here neglected. In addition one would have to use the full covariance
matrix of the BOOMERanG band powers rather than their quoted error
bars only. However, we believe that the result shown on
Fig.~\ref{fig1} provides an indication that the role of topological
defects in structure formation should not be underestimated just on
the ground that they happen to be unable to reproduce all the data if
they are the sole source of primordial fluctuations.

\vskip 0.5cm

It is a pleasure to thank B.~Carter, N.~Deruelle, R.~Juszkiewicz,
A.D.~Linde and J.-P.~Uzan for enlightening comments.

\end{document}